\documentclass[graphicx,amssymb,amsmath,reprint]{revtex4-1}

\draft 

\begin{document}


\title{Nonlinear charge transport in bipolar semiconductors due to electron
    heating} 



\author{S. Molina-Valdovinos}
\email{sergiom@fisica.uaz.edu.mx}
\affiliation{Universidad Aut\'onoma de Zacatecas, Unidad Acad\'emica de F\'isica,
Calzada Solidaridad esq. Paseo, La Bufa s/n, CP 98060, Zacatecas, Zac. M\'exico.}

\author{Yu. G. Gurevich}
\affiliation{Centro de Investigaci\'on y de Estudios Avanzados del IPN, Departamento de F{\'i}sica, Av. IPN 2508, M\'exico D.F., CP 07360, M\'exico.}


\date{\today}

\begin{abstract}
It is known that when strong electric field is applied to a semiconductor sample, the current voltage characteristic deviates from the linear response. In this letter, we propose a new point of view of nonlinearity in semiconductors which is associated with the electron temperature dependence on the recombination rate. The heating of the charge carriers breaks the balance between generation and recombination, giving rise to nonequilibrium charge carriers concentration and nonlinearity.
\end{abstract}


\maketitle 


The study of nonlinear transport deals with physical processes that arise when a sufficiently strong electric field is applied to a semiconductor sample, so that the current deviates from the linear response. 

If we have a metal-semiconductor (Ohmic) contact, in an external electric field, the electrons move inside of a semiconductor changing the concentration of carriers in the volume (injection of majority carriers). The current-voltage characteristic (CVC) changes from the linear to quadratic and then back to linear. This effect is known as Mott's law\cite{NFM}. In the Schottky barrier\cite{WS,BLS}, when the equilibrium concentration of charge carriers near the contact is less than the volume concentration, an applied external electric field causes a diode type CVC. Another mechanism that leads to nonlinearity is the minority carrier injection in p-n contact\citep{CTS,SZE}. As a consequence of the Peltier and Thompson effect, a mechanism of nonlinearty arises, when an electric field is applied near the contact. We have heating or cooling which generates a nonlinear electrical current in the semiconductor\cite{GNL}. All these mechanisms of nonlinearity of CVC are due to contacts.

There are other mechanisms of nonlinearity due to the volume.

In bipolar semiconductors, nonequilibrium charge carriers is a mechanism that generates nonlinearity in the density of the current\cite{MML}. Previous calculations have been addressed to nonlinearity caused by an impact ionization\cite{LVK,SZE}, carrier lifetime changes \cite{MA}, intervalley redistributions of carriers \cite{MHJ}.

In semiconductors there are mechanisms of nonlinearity in the density of current associated with mobility (hot electrons). One of this mechanism is due to the effective mass that depends on the energy, $m^{\ast}=m^{\ast}(E)$\cite{KEO}. Another mechanism is due to the dependence of the momentum relaxation time on energy\cite{EC,FPR,LR}.

As a field increases, the average energy of the electrons and holes increases, too (hot carriers). Therefore the system is out of equilibrium, i.e., when the effective temperature for electrons $T_{n}$  and holes $T_{p}$ are greater than the lattice temperature $T_{ph}$. As a result, electrons and holes tend to occupy higher quantum states in the conduction and valence bands, respectively. The balance between thermal generation and recombination of electrons and holes is broken\cite{EC,EMC,JZ}. The dependence of the rate of recombination on the temperature of carriers generates a change in the carrier concentration\cite{INVYu}. So, the non equilibrium concentration for electrons (holes) is connected with the nonequilibrium temperature of carriers which changes the electrical conductivity. Experimental verification of the considered effect is shown in Ref. \onlinecite{EMCJZ,JZEMC}.

However, due to the assumption made that the population of the impurity level does not depend on heating, the results turn out to be incorrect if the heating of electrons and holes is not equal (electron and holes temperatures are unequal). However, the question of when this effect takes place was left open. In spite of all these facts, the considered mechanism of nonlinearity has not received further development.

In general, in semiconductors the effective mass of holes is greater than the effective mass of electrons $(m_{p}^{*}>m_{n}^{*})$, and the electron-phonon interaction is more quasielastic, it implies that electrons acquire more energy (hot electrons), and their temperature increases above the phonon temperature, i.e., $T_{e}>T_{ph}$. By contrast, the hole-phonon interaction is less quasielastic. The holes do not obtain energy (not heated), so that the temperature of holes is equal to phonon temperature $(T_{p}\approx T_{ph} \approx T_{0})$. It is important to emphasize that the heating only of electrons leads to the appearance of both the nonequilibrium electrons and nonequilibrium holes.

For simplicity we study the case when the electric field is low, in this case we do not have hot electrons but rather a warm electron condition\cite{YuGG2006,YGC2007}. Unfortunately, these publications did not have a clear statement of the problem. So, analyzing the results obtained is very difficult. There were two more articles related to this problem. In publication \onlinecite{MCI} the statement of the problem was incorrect, and in \onlinecite{GGC} only intrinsic semiconductor thin films were considered. Under warm electron conditions, the electron temperature $T_{n}=T_{0}+\delta T_{n}$ with $\delta T_{n}\ll T_{0}$. The concentration for electrons and holes are  $n=n_{0}+\delta n$, $p=p_{0}+\delta p$, $\delta n \ll n_{0}$, $\delta p\ll p_{0}$.

As was shown in Ref. \onlinecite{INVGNL}, the recombination rate depends on the electron temperature. In the case of band-band recombination\cite{IL}, we have:  
\begin{equation}\label{E1}
R=\frac{1}{\tau_{bb}} \frac{n_{0}p_{0}}{n_{0}+p_{0}} \left(\frac{\delta n}{n_{0}} + \frac{\delta p}{p_{0}} - \frac{1}{\alpha}\left.\frac{\partial \alpha}{\partial T_{n}}\right|_{T_{n}=T_{0}}\delta T_{n}\right),
\end{equation}
\noindent{and $\tau_{bb}$ is the effective life time of electrons-hole pair of band-band recombination, defined as:}
\begin{equation}\label{E2}
\tau_{bb}=\left[\alpha(n_{0}+p_{0})\right]^{-1},
\end{equation}

\noindent{with $\alpha$ the capture coefficient of electrons by holes.}

The recombination rate through the traps can be written as follows\cite{INVGNL}:
\begin{eqnarray}\label{E3}
R=\frac{1}{\tau_{t}}\frac{n_{0}p_{0}}{n_{0}+p_{0}}\left[ \frac{\delta n}{n_{0}} + \frac{\delta p}{p_{0}} + r \delta T_{\rm n}\right],
\end{eqnarray}

\noindent{where}
\begin{eqnarray} \label{E4}
r=\frac{n_{1}^{0}}{p_{0}} \frac{1}{\alpha_{\rm n}(T_{0})} \left.\frac{\partial \alpha_{\rm n}}{\partial T_{\rm n}}\right|_{T_{\rm n}=T_{0}}
\end{eqnarray}

\noindent{and $\tau_{t}$ is the effective life time of electrons-hole pair of doped semiconductors.}
\begin{equation}\label{E5}
\tau_{t}=\left[\frac{\alpha_{\rm n}(T_{0})\alpha_{\rm p}(T_{0})N_{t}(n_{0}+p_{0})}{\alpha_{\rm n}(T_{0})(n+n_{1}^{0})+\alpha_{\rm p}(T_{0})(p+p_{1}^{0})}\right]^{-1},
\end{equation}
 
\noindent{here, $N_{t}$ is the impurity concentrations, $\alpha_{\rm n}(T_{0},T_{n})$ and $\alpha_{\rm p}(T_{0})$ are the electron and hole capture coefficients, $n_{1}^{0}$ $(p_{1}^{0})$ the electron (hole) concentration when the Fermi level matches the activation energy of the impurity in equilibrium, $n_{1}^{0}=\nu_{n}(T_{0})\exp{[-\varepsilon_{t}/T_{0}]}$; $p_{1}^{0}=\nu_{p}(T_{0})\exp{[\varepsilon_{t}-\varepsilon_{g}/T_{0}]}$, $\varepsilon_{t}$ is the impurity energy level, and $\nu_{n}(T_{0})=1/4(2m_{n}T_{0}/h\pi^{2})^{3/2}$, $\nu_{p}(T_{0})=1/4(2m_{n}T_{0}/h\pi^{2})^{3/2}$ are the densities of state at the bottom of the conduction band and top of the valence band.}

If the sample has a finite size along z-axis and the electron system is thermally insulated at the surfaces (adiabatic system), there is not a mechanism for energy relaxation of carriers at the surfaces. If, furthermore, the mechanisms of surface recombination are absent, the density of currents for electrons and holes over the surfaces of the samples is null, $j_{\rm nz}\vert_{z=\pm b}=j_{\rm pz}\vert_{z=\pm b} = 0$. In the same way the electron heat flux is null over the surfaces\cite{FGB}, $Q_{n}\vert_{z=\pm b}=0$. Under this condition, all parameters of the bipolar semiconductors do not depend on the coordinates. 

From the continuity equations\cite{SZE}, $\nabla \cdot j_{n}= R$ and $\nabla \cdot j_{p}=- R$ we have $\nabla \cdot j_{n}=\nabla \cdot j_{p}=0$. It implies that the volumetric recombination is zero, $R=0$, but the effective life time of electron-hole recombination is nonzero, i.e., $\tau_{bb} \neq 0$ and $\tau_{t} \neq 0$. Another consequence of this consideration is that it does not depend on whether we have weak recombination or strong recombination. The result is the same. Under the condition $R=0$, we have for band-band recombination
\begin{equation}\label{E6}
\frac{\delta n}{n_{0}} + \frac{\delta p}{p_{0}} - \frac{1}{\alpha}\left.\frac{\partial \alpha}{\partial T_{n}}\right|_{T_{n}=T_{0}} \delta T_{n}=0,
\end{equation}

\noindent{and for the recombination rate through the traps}
\begin{eqnarray}\label{E7}
\frac{\delta n}{n_{0}} + \frac{\delta p}{p_{0}} + r \delta T_{\rm n}=0.
\end{eqnarray}

Note, that the electric field $E_{x}$ which was applied along the x-axis, is constant along all the sample. From the Poisson equation\cite{KS}, we have electroneutrality condition, $\rho=\rho_{0}+\delta \rho = 0$. 

From condition $\delta \rho \approx 0$, and if the mechanism of recombination is band-band, then $\delta n \approx \delta p$. 

In the case of a semiconductor that contains an impurity concentration\cite{INVGNL} $N_{t}$ and under the electroneutrality condition, the nonequilibrium charge carrier concentrations $\delta n$, $\delta p$ and $\delta T_{n}$, are related by\cite{OYT},
\begin{eqnarray}\label{E8}
\delta p= \zeta_{1}\delta n +\zeta_{2} \delta T_{\rm n},
\end{eqnarray}

\noindent{where}  
\begin{eqnarray*}\label{E9}
\zeta_{1}&=&\frac{\alpha_{\rm n}(N_{t}n_{0}+\frac{n_{\rm t0}^{2}n_{1}^{0}}{n_{0}})+\alpha_{\rm p}N_{t}p_{1}^{0}}{\alpha_{\rm n}N_{t}n_{0}+\alpha_{\rm p}(N_{t}p_{1}^{0}+n_{\rm t0}^{2})},\\
\zeta_{2}&=&\left[\frac{n_{1}^{0}n_{\rm t0}^{2}}{\alpha_{\rm n}N_{t}n_{0}+\alpha_{\rm p}(N_{t}p_{1}^{0}+n_{\rm t0}^{2})}\right]\frac{\partial \alpha_{\rm n}(T_{0})}{\partial T_{\rm n}},
\end{eqnarray*}

\noindent{where $n_{t0}$ is the equilibrium concentration of electrons located on the impurity levels.}

\noindent{The energy balance equation for electrons is\cite{FGB},}
\begin{equation}\label{Eq10}
n \nu_{n\varepsilon}(T_{n})(T_{n}-T_{0})=\vec{j}_{n}\cdot\vec{E},
\end{equation}

\noindent{where $n \nu_{n\varepsilon}(T_{n})(T_{n}-T)$ describes the intensity of electron-phonon energy exchange with $\nu_{n\varepsilon}$ the electron energy relaxation frequency, $\vec{E}=(E_{x},0,0)$ the external electric field, $\vec{j}_{n}\cdot\vec{E}$ is the Joule effect.}  

From the energy balance equation for electrons we obtain the nonequilibrium temperature for electrons 
\begin{equation}\label{E11}
\delta T_{n}=\frac{\sigma_{xx}^{n0}E^{2}_{x}}{n_{0}\nu_{n\varepsilon}(T_{0})},
\end{equation}

\noindent{where $\sigma_{xx}^{n0}=n_{0}e^{2}\tau_{n}(T_{0})/m_{n}$, and $\tau_{n}(T_{n})=\tau_{0}(T_{n}/T_{0})^{q_{n}}$. The exponent $q_{n}$ take different values depending of the mechanism of  dispersion (see for example \onlinecite{FPR,VFG}).}
 
In the case of band-band recombination we have the nonequilibrium concentration for electrons and holes are equal, $\delta n \approx \delta p$. From the equation (\ref{E6}) the relation between the nonequilibrium concentrations and the temperature is 
\begin{eqnarray}\label{E12}
\delta n&=&\delta p= \frac{n_{0}p_{0}}{n_{0}+p_{0}}\left[\left.\frac{\partial \ln \alpha}{\partial T_{n}}\right|_{T_{n}=T_{0}}\right]\delta T_{n}\nonumber\\ 
&=&\frac{n_{0}p_{0}}{n_{0}+p_{0}}\left[\left.\frac{\partial \ln \alpha}{\partial T_{n}}\right|_{T_{n}=T_{0}}\right]\frac{\sigma_{xx}^{n0}E^{2}_{x}}{n_{0}\nu_{\varepsilon}(T_{0})}.
\end{eqnarray}

From equations (\ref{E7}) and (\ref{E8}) (the recombination rate through the traps), we have that the relation between the nonequilibrium concentration for electrons and holes with the temperature is:
\begin{subequations}
\begin{eqnarray}
\delta n&=&-\left( \frac{n_{0}p_{0}r+n_{0}\zeta_{2}}{p_{0}+n_{0}\zeta_{1}}\right)\delta T_{n},\\
\delta p&=&\left( \frac{p_{0}\zeta_{2}-n_{0}p_{0}r\zeta_{1}}{p_{0}+n_{0}\zeta_{1}}\right)\delta T_{n}.
\end{eqnarray}
\end{subequations}

The nonlinear density of current along the x-axis takes the form\cite{YGC2007,YuGC2013},
\begin{eqnarray}\label{E14}
j_{x}&=&j_{nx}+j_{px}\\
&=&\sigma_{xx}^{n0}\left(1+\frac{q_{n}}{T_{0}}\delta T_{n}+\frac{\delta n}{n_{0}} \right)E_{x}+\sigma_{xx}^{p0}\left(1+\frac{\delta p}{p_{0}} \right)E_{x},\nonumber
\end{eqnarray}  

\noindent{where $\sigma_{xx}^{p0}=p_{0}e^{2}\tau_{p}(T_{0})/m_{p}$.}
The excess of nonequilibrium carriers generated by the heating of electrons generates a new contribution to the CVC. This new contribution comes from the dependence of the concentration of electrons and holes on the temperature.

In the case of band-band recombination, the nonlinear density of current along the x-axis takes the form (see equation (\ref{E12})),
\begin{equation}\label{E15}
j_{x}=j_{nx}+j_{px}=\sigma_{0}\left(1+\gamma_{0}E_{x}^{2}\right)E_{x}
\end{equation}

\noindent{with}
\begin{eqnarray*}\label{E16}
\sigma_{0}&=&\sigma_{xx}^{n0}+\sigma_{xx}^{p0},\\
\gamma_{0}&=&\frac{1}{\sigma_{0}}\left[\frac{q_{n}\sigma_{xx}^{n0}}{T_{0}}+\right.\\
&+&\left.\left[\frac{\sigma_{xx}^{n0}}{n_{0}}+\frac{\sigma_{xx}^{p0}}{p_{0}}\right]\frac{1}{\alpha}\left.\frac{\partial \alpha(T_{n})}{\partial T_{n}}\right|_{T_{n}=T_{0}} \frac{n_{0}p_{0}}{n_{0}+p_{0}}\right]\frac{\sigma_{xx}^{n0}}{n_{0}\nu_{\varepsilon}(T_{0})}.
\end{eqnarray*}

Note that the classical theory of CVC appears when $\gamma_{0}=0$. The nonlinearity comes from the heating of electrons and recombination processes. The second term ($\gamma_{0}$) in equation (\ref{E15}) connects with appearance of nonequilibrium electrons and holes. Note that if we consider the case of intrinsic semiconductor ($n_{0}= p_{0}$) and if $q_{n}=0$ then $\gamma_{0}=(\left.\frac{\partial \ln \alpha}{\partial T_{n}}\right|_{T_{n}=T_{0}})\frac{e^{2}\tau_{0}}{2m_{n}\nu_{\varepsilon}(T_{0})}$. The appearance of nonlinearity of CVC is maintained even in absence of a mechanism of dispersion.  

When we have the recombination rate through the traps, the nonlinear density of current along the x-axis takes the form,
\begin{equation}\label{E17}
j_{x}=j_{nx}+j_{px}=\sigma_{0}\left(1+\gamma_{1}E_{x}^{2}\right)E_{x},
\end{equation}

\noindent{with}
\begin{eqnarray*}\label{E18}
\gamma_{1}=&\frac{1}{\sigma_{0}}&\left[\frac{q_{n}\sigma_{xx}^{n0}}{T_{0}}-\frac{\sigma_{xx}^{n0}}{n_{0}}\left( \frac{n_{0}p_{0}r+n_{0}\zeta_{2}}{p_{0}+n_{0}\zeta_{1}}\right)+\right. \\
&+&\left. \frac{\sigma_{xx}^{p0}}{p_{0}}\left( \frac{p_{0}\zeta_{2}-n_{0}p_{0}r\zeta_{1}}{p_{0}+n_{0}\zeta_{1}}\right)\right]\frac{\sigma_{xx}^{n0}}{n_{0}\nu_{\varepsilon}(T_{0})}.
\end{eqnarray*}

From $\gamma_{1}$ is evident that the nonlinearity of CVC depends on the recombination process. In this case the nonlinearity comes from the nonlinear concentrations of electrons and holes. 

In conclusion, the heating of electrons generated by the external electric field  causes a break between the thermal generation and recombination processes. This break gives rise to nonequilibrium charge carriers. The change in the nonequilibrium concentrations generates a change in the CVC from the linear to nonlinear.\\


The Authors wish to thank CONACYT-M\'exico for partial financial support. The Author S. Molina-Valdovinos wishes to thank PROMEP-M\'exico(UAZ-PTC-185) for partial financial support.


%
%

%


\bibliographystyle{unsrt}
\bibliography{TEWNMF}

\begin{thebibliography}{10}

\bibitem{NFM}
N.F. Mott.
\newblock {\em Proc. Camb. Phil. Soc.}, 34:568--572, 1938.

\bibitem{WS}
W.~Schottky.
\newblock {\em Naturwissenschaften}, 26:834, 1938.

\bibitem{BLS}
B.~L. Sharma.
\newblock {\em Metal-semiconductor Schottky barrier junctions and their
  applications}.
\newblock Plenum Press. New York, 1984.

\bibitem{CTS}
C.~T. Sah, R.~N. Noyce, and W.~Shockley.
\newblock {\em Proc. IRE}, 45:1228, 1957.

\bibitem{SZE}
S.~M. Sze.
\newblock {\em Physics of semiconductors devices}.
\newblock New York: Wiley, 1981.

\bibitem{GNL}
G.~N. Logvinov, J.~E. Vel\'azquez, I.~M. Lashkevich, and Yu.~G. Gurevich.
\newblock {\em Appl. Phys. Lett.}, 89:092118, 2006.

\bibitem{MML}
Yuri G.~Gurevich y~Miguel Mel\'endez~Lira.
\newblock {\em Fenómenos de contacto y sus aplicaciones en celdas solares}.
\newblock Fondo de cultura econ\'omica, M\'exico, 2010.

\bibitem{LVK}
L.V. Keldish.
\newblock {\em Sov. Phys.—JETP}, 21:1135, 1965.

\bibitem{MA}
M.~Asche, H.~Kostial, and O.G. Sarbey.
\newblock {\em Phys. Status Solidi b}, 91:521, 1979.

\bibitem{MHJ}
M.H. Jorgensen.
\newblock {\em Phys. Rev. B}, 18:5657, 1978.

\bibitem{KEO}
E.~O. Kane.
\newblock {\em J. Chem. Phys. Solids}, 1:249, 1957.

\bibitem{EC}
Esther Conwell.
\newblock {\em High field transport in semiconductors}.
\newblock Academic press, New York, 1967.

\bibitem{FPR}
Yu.~G. Gurevich and F.~P\'erez Rodr\'iguez.
\newblock {\em Fen\'omenos de Transporte en Semiconductores}.
\newblock Fondo de Cultura Economica, M\'exico, 2007.

\bibitem{LR}
L.~Reggiani.
\newblock {\em Hot electron transport in Semiconductors}.
\newblock Springer-Verlag, Berlin Heidelberg, 1985.

\bibitem{EMC}
E.~M. Conwell.
\newblock {\em J. Phys. Chem. Solids}, 17:342, 1961.

\bibitem{JZ}
J.~Zucker and E.~M. Conwell.
\newblock {\em J. Phys. Chem. Solids}, 22:149, 1961.

\bibitem{INVYu}
Yu.~G. Gurevich and I.~N. Volovivhev.
\newblock {\em Physical Review B}, 60:7715, 1999.

\bibitem{EMCJZ}
E.~M. Conwell and J.~Zucker.
\newblock {\em J. Phys. Chem. Solids}, 22:141, 1961.

\bibitem{JZEMC}
J.~Zucker and E.~M. Conwell.
\newblock {\em J. Phys. Chem. Solids}, 23:1549, 1962.

\bibitem{YuGG2006}
Yu.~G. Gurevich and G.~Gonzalez de~la Cruz.
\newblock {\em Semicond. Sci. Technol.}, 21:1686--1690, 2006.

\bibitem{YGC2007}
G.~Gonzalez de~la Cruz and Yu.~G. Gurevich.
\newblock {\em J. Phys.: Condens. Matter}, 19:456220, 2007.

\bibitem{MCI}
Yu.~G. Gurevich, H.~Lohvinov, M.~Cruz-Irisson, O.~Titov, G.Espejo-L\'opez, and
  I.~Volovichev.
\newblock {\em Phys. Stat. Sol. (c)}, 1:S100--S103, 2004.

\bibitem{GGC}
Yu. G.~Gurevich G.~Gonzalez de~la Cruz.
\newblock {\em Rev. M\'ex. de Fis.}, 56:211—216, 2010.

\bibitem{INVGNL}
I.~N. Volovichev, G.~N. Logvinov, O.~Yu. Titov, and Yu.~G. Gurevich.
\newblock {\em J. Appl. Physics}, 95:4494, 2004.

\bibitem{IL}
I.~Lashkevych and Yu.~G. Gurevich.
\newblock {\em International Journal of Heat and Mass Transfer}, 92:430--434,
  2016.

\bibitem{FGB}
F.~G. Bass, V.~S. Bochkov, and Yu.~G. Gurevich.
\newblock {\em Soviet Physics-Semiconductors}, 7:3--32, 1973.

\bibitem{KS}
K.~Seeger.
\newblock {\em Semiconductors Physics}.
\newblock Springer, Berlin, 1985.

\bibitem{OYT}
Yu.~G. Gurevich, J.~E. Vel\'azquez-P\'erez, G.~Espejo-L\'opez, I.~N.
  Volovichev, and O.~Yu. Titov.
\newblock {\em J. of Appl. Phys.}, 101:023705, 2007.

\bibitem{VFG}
V.~F. Gantmakher and Y.~B. Levinson.
\newblock {\em Carrier sacatering in metals and semiconductors}.
\newblock North-Holland, Amsterdam, 1987.

\bibitem{YuGC2013}
G.~Gonzalez de~la Cruz and Yu.~G. Gurevich.
\newblock {\em J. of Appl. Phys.}, 113:023504, 2013.

\end{thebibliography}

\end{document}